# A comparative *ab-initio* investigation of the physical properties of cubic Laves phase compounds $X$Bi$_2$ ($X$ = K, Rb)

Jahid Hassan, M. A. Masum, S. H. Naqib*
Department of physics, University of Rajshahi, Rajshahi 6205, Bangladesh
Corresponding author email: salehnaqib@yahoo.com

**Abstract**

In this study, we looked into a number of physical properties of alkali-bismuth compounds $X$Bi$_2$ ($X$ = K, Rb), in the cubic Laves phase (symmetry Fd$\bar{3}$m) using the density functional theory (DFT). The structural, elastic behavior along with Pugh's ratio, Poisson's ratio, Cauchy pressure, anisotropy indices, micro- and macro-hardness, thermo-physical properties such as Debye temperature, sound velocities, Grüneisen parameter, and melting temperature, electronic band structure, and optoelectronic properties has been explored. The computed ground-state lattice parameters and unit cell volume are in close accordance with the known theoretical and experimental findings. The elastic, thermo-physical, and optoelectronic properties of $X$Bi$_2$ ($X$ = K, Rb) are investigated for the first time in this study. The computed elastic constants satisfied the mechanical stability criteria. The estimated Pugh's ratio, Poisson's ratio, and Cauchy pressure signify the ductility of the compounds. In order to understand the electronic properties, band structures and electronic energy densities of states have been explored. These compounds exhibit metallic characteristics in their electrical band structures. We have done a complete investigation on the reflectivity, absorption coefficient, refractive index, dielectric function, optical conductivity, and loss function of these metals. These compounds possess a low Debye temperature, thermal conductivity, and melting point. The optical absorption, reflectivity spectra and the refractive index of $X$Bi$_2$ ($X$ = K, Rb) show that they can be used as solar reflector and ultraviolet absorber. The majority of the findings in this study are novel.

## 1. Introduction

Laves phase are ideal for studying the fundamentals of intermetallic phases due to their large representative number, polytypism, extended homogeneity ranges, and simple crystal structure [1–3]. Laves phase compounds play a significant role in various functional and structural applications, including hydrogen storage material (nickel-metal hydride batteries), magneto-mechanical sensors, and wear- and corrosion-resistant coatings [1,4]. They are also used in high temperature environment such as aerospace, power generation, high-temperature steel creep-strengthening, new alloy designing, and other structural materials [5,6]. Moreover, they are also applicable for using in solar cells and optoelectronic devices, due to their high efficiency, low toxicity, and good environmental stability [5,7].

The chiral phonon modes emerge in the cubic Laves phase $X$Bi$_2$ ($X$ = K, Rb) compounds due to its geometry of the crystal structure, which are associated with the circulation of atoms around their equilibrium positions [8,9]. It possesses both time reversal symmetry and inversion symmetry and results in chiral phonons with vanishing total phonon circular polarization (PCP). It is possible for chiral phonons with a non-zero total phonon circular polarization (PCP) to be hosted in a doped system, which opens up new possibilities for chiral phonon engineering and experimental investigation [10]. In addition, the chiral modes for the wave vectors at the boundary of the Brillouin zone are realized and exhibit two- and three-fold rotational symmetry. In addition to the chirality in the phonon spectrum in some of the Laves phase compounds, they also exhibit a set of physical properties suitable for electronic, optical, mechanical, and thermal applications [11-14].

The face-centered cubic Laves phase compounds $X\text{Bi}_2$ ($X$ = K, Rb), with $Fd\bar{3}m$ space group (no. 227) symmetry, is isostructural to the cubic $\text{MgCu}_2$ Laves phase (C15-type) [1,11–14]. 8 $X$ ($X$ = K, Rb) and 16 Bi atoms make up each conventional unit cell in the compound. Tetrahedra made of four Bi atoms can share their vertices to form a three-dimensional network connected to one another. As a result, the network of Bi tetrahedra is entangled with the arrangement of $X$ ($X$ = K, Rb) atoms in a diamond sublattice. There is a hyperkagome structure created by the Bi sublattice. Electronic structure computations reveal the electronic states of these systems, with the Bi-Bi bond being less metallic than the $X$-Bi ($X$ = K, Rb) bond [15,16]. These bonding features largely control the elastic and mechanical properties of these compounds.

Due to its interesting characteristics and possible uses in various functional, structural applications and also in the field of superconductivity, we have chosen two face-centered cubic systems in the Laves phase (C15-type), $X\text{Bi}_2$ ($X$ = K, Rb) to examine their physical properties in this work. These compounds are superconductors exhibiting type-I behavior. The compound $\text{KBi}_2$ with $T_C$ = 3.57 K, and $\text{RbBi}_2$ with $T_C$ = 4.15 K has gathered considerable interest over time [16]. The structural, electronic, phonon dynamics, magnetic, and superconducting state properties of $X\text{Bi}_2$ ($X$ = K, Rb) have been investigated in previous researches [16,17].

Additionally, the caloric characteristics, magnetic field dependency of the resistivity, and pressure-dependent XRD patterns were all studied [18]. Over a broad pressure range, the structural and superconducting characteristics of $X\text{Bi}_2$ ($X$ = K, Rb) were studied. A departure from the s-wave dirty limit model for the Cooper pair symmetry was suggested by the temperature dependence of the lower critical magnetic field of $\text{RbBi}_2$ under pressure [16,17].

But there are still a lot of significant physical characteristics of $X\text{Bi}_2$ ($X$ = K, Rb) that remain unknown. The majority of physical characteristics important to future applications, such as elastic behavior, mechanical anisotropy, thermo-physical features, and optical characteristics have not been thoroughly studied so far. There are still many bonding related topics that need to be researched, such as the analysis of Cauchy pressure, Pugh's ratio, Poisson's ratio, micro- and macro-hardness, machinability index, Kleinman parameter, anisotropy in elastic moduli, and many more. There are several thermo-physical parameters that have not yet been completely studied, including the Debye temperature, melting temperature, lattice thermal conductivity, sound velocities, thermal expansion coefficient, heat capacity, dominating phonon wavelength, and minimum thermal conductivity. Thermo-physical parameters give information on how a material behaves at various temperatures. We have looked at these properties in this study and comparisons have been made. It is crucial to understand optical qualities when choosing a system for applications in optoelectronic devices. Therefore, optical constants with incident photon energy have been studied for the chosen compounds.

The information given above makes it evident that there is still much to learn about the physical characteristics of these binary Laves phase materials. It is therefore, of scientific interest to look at these unexplored aspects. Due to a paucity of basic information, we used *ab-initio* calculations to address the structural, elastic, electronic, optical, and thermo-physical characteristics of $X$Bi$_2$ ($X$ = K, Rb) compounds in this paper.

The remainder of this work is structured as follows: In section 2, a brief explanation of the computational approach used in this study is given. In section 3, we provide the computational findings and evaluations. In section 4, the key features are discussed together with the relevant findings for this investigation.

## 2. Computational methodology

The most popular framework for *ab-initio* calculations on crystalline materials is the density functional theory (DFT), which finds a crystalline system's ground state by solving the Kohn-Sham equation [19] with periodic boundary conditions (including Bloch states). The physical characteristics of the compounds were investigated using the DFT-based CAmbridge Serial Total Energy Package (CASTEP) [20]. This code employs a technique known as the total energy plane-wave pseudopotential. This investigation utilized local density approximation (LDA) [21,22], generalized gradient approximations (GGA) of the Perdew-Burke-Ernzerhof (PBE) [23], GGA of the Perdew-W91 [24], GGA of the RPBE [25], and GGA of the PBEsol [23] exchange-correlation functionals. GGA of the RPBE delivers better ground state structural parameters in comparison to the experimental lattice constants. The significance of accurate geometry optimization is crucial. Since GGA of the RPBE produces the best structural characteristics for the $X$Bi$_2$ ($X$ = K, Rb) compounds, we have reported results for this in the following sections. In order to calculate the electron-ion interactions, ultra-soft Vanderbilt type pseudopotentials were used [26]. The norm-conserving constraint is loosened in this method generating a smooth and computation-friendly pseudopotential that drastically cuts calculation time without sacrificing accuracy.

The valence electron configurations of K, Rb, and Bi atoms are taken as [$3s^2\ 3p^6\ 4s^1$], [$4s^2 4p^6 5s^1$], and [$6s^2 6p^3$], respectively. In this study, the first irreducible Brillouin zone (BZ) k-point sampling is performed utilizing the Monkhorst-Pack scheme of size 6 × 6 × 6, and the cut-off energy for the plane wave expansion was taken to be 400 eV for both $X$Bi$_2$ ($X$ = K, Rb) compounds. In order to perform the specific k-point sampling of the Brillouin zone (BZ), the Monkhorst-Pack technique is widely used [27]. For further information on k-point grids and pseudopotentials, please see Ref. [28].

The single crystal elastic constants ($C_{ij}$) of the cubic structure can be obtained using the stress-strain technique [29]. The elastic constants $C_{11}$, $C_{12}$, and $C_{44}$ are all unique to cubic crystals. By applying the Voigt-Reuss-Hill (VRH) method [30,31], it is possible to compute all the elastic properties, including the bulk modulus ($B$), shear modulus ($G$), and the Young's modulus based on the single crystal elastic constants $C_{ij}$. Using the outcomes of the optimized geometry of $X$Bi$_2$ ($X$ = K, Rb), the total electronic energy density of states (TDOS), partial density of states (PDOS), and electronic band structures are estimated.

The complex dielectric function, $\varepsilon(\omega) = \varepsilon_1(\omega) + i\varepsilon_2(\omega)$, can be used to get all the optical constants that rely on energy or frequency of the electromagnetic wave. The well-known Kramers-Kronig equation's [32] imaginary part is used to calculate the real portion of the dielectric function. The momentum representation of the matrix elements of the photon-induced transition between occupied and unoccupied electronic states can be used to calculate the imaginary part, $\varepsilon_2(\omega)$, of the complex dielectric function. The imaginary component of the dielectric function has the following CASTEP-contained expression:

$$\varepsilon_2(\omega) = \frac{2e^2\pi}{\Omega\varepsilon_0} \sum_{k,v,c} |\langle \psi_k^c | \hat{u}.\vec{r} | \psi_k^v \rangle|^2 \delta(E_k^c - E_k^v - E) \quad (1)$$

In the above expression, $\Omega$ is the volume of the unit cell, $\omega$ is the angular frequency (or equivalently energy) of the incident electromagnetic wave (photon), $\hat{u}$ is the unit vector defining the polarization direction of the incident electric field, $e$ is the electronic charge, $\psi_k^c$ and $\psi_k^v$ are the conduction and valence band wave functions at a given wave-vector k, respectively. The conservation of energy and momentum throughout the optical transition is implemented by the delta function in Equation (1). Once the dielectric function $\varepsilon(\omega)$ is known at various energies, all other significant optical properties may be deduced from it, using the interrelations [33–35] shown below. The following relations can be used to compute the real, $n(\omega)$, and imaginary, $k(\omega)$, components of the complex refractive index:

$$n(\omega) = \frac{1}{\sqrt{2}} [\{\varepsilon_1(\omega)^2 + \varepsilon_2(\omega)^2\}^{1/2} + \varepsilon_1(\omega)]^{1/2} \quad (2)$$

$$k(\omega) = \frac{1}{\sqrt{2}} [\{\varepsilon_1(\omega)^2 + \varepsilon_2(\omega)^2\}^{1/2} - \varepsilon_1(\omega)]^{1/2} \quad (3)$$

Again, using the complex refractive index components, the reflectivity, $R(\omega)$, may be calculated:

$$R(\omega) = \left|\frac{\tilde{n}-1}{\tilde{n}+1}\right| = \frac{(n-1)^2 + k^2}{(n+1)^2 + k^2} \quad (4)$$

Furthermore, the following equations can be used to calculate the absorption coefficient, $\alpha(\omega)$, optical conductivity, $\sigma(\omega)$, and energy loss function, $L(\omega)$:

$$\alpha(\omega) = \frac{4\pi k(\omega)}{\lambda} \quad (5)$$

$$\sigma(\omega) = \frac{2W_{cv}\hbar\omega}{\vec{E}_0^2} \quad (6)$$

$$L(\omega) = \operatorname{Im}\left[-\frac{1}{\varepsilon(\omega)}\right] \quad (7)$$

In equation (6), $W_{cv}$ is the optical transition probability per unit time.

The Debye temperature has been calculated using the acoustic velocities in the crystals. The crystal density, elastic constants, and elastic moduli of $X$Bi$_2$ ($X$ = K, Rb) are used to compute the other thermo-physical parameters. The formalisms described above have been used previously to a large number of condensed phases including technologically prominent MAB and MAX phases [36-41] with high degree of accuracy.

### 3. Results and discussions
3.1. Structural properties

The symmetry and crystal structure of a substance have a significant impact on a number of physical properties, such as elastic constants, electrical band structure, and optical characteristics. Every aspect of a solid's physical and electrical properties is governed by the arrangement of its atoms, their distances from one another, and electronic states. The crystal structure of $X\text{Bi}_2$ ($X$ = K, Rb) compounds is cubic based on the Laves phase (C15-type) with space group $Fd\bar{3}m$ (no. 227) [1,10]. An illustration of the crystal structure of $X\text{Bi}_2$ ($X$ = K, Rb) compounds is represented below in Figure 1. A single unit cell of $X\text{Bi}_2$ ($X$ = K, Rb) consists 8 $X$ ($X$ = K, Rb) atoms and 16 Bi atoms; 8 formula units. There are only two atomic positions in a unit cell: $X$ ($X$ = K, Rb) atoms at (1/4, 3/4, 1/4) and Bi atoms at (3/8, 1/8, 3/8).

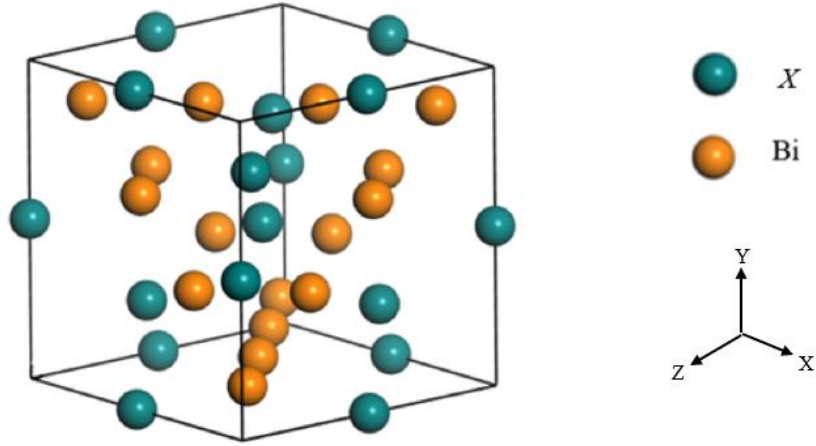

**Figure 1:** Schematic crystal structure of $X\text{Bi}_2$ ($X$ = K, Rb) compounds. The crystallographic directions are also shown.

The lattice parameters of the relevant compounds are totally relaxed during the geometry optimization. The angles of the lattice are: $\alpha = \beta = \gamma = 90°$. The lattice parameters of $\text{RbBi}_2$ are marginally greater than that of $\text{KBi}_2$. The difference in the atomic radii of K, and Rb can be partly responsible for this variation. Table 1 compares the lattice parameters derived through geometry optimization with existing theoretical data.

**Table 1**: The optimized lattice parameters $a$ (Å), and optimized cell volume $V$ (Å$^3$) of $X\text{Bi}_2$ ($X$ = K, Rb) compounds.

| Compound | $a$ | $V$ | Functionals | Ref. |
|---|---|---|---|---|
| | 9.52 | 863.98 | GGA(RPBE) | This work |
| | 9.66 | 902.54 | GGA(PBE) | [10][Theo.] |
| $\text{KBi}_2$ | 9.25 | 791.71 | ---- | [4][Theo.] |
| | 9.52 | 863.61 | ---- | [42][Theo.] |

| | 9.56 | 898.63 | ---- | [17][Theo.] |
| --- | --- | --- | --- | --- |
| | 9.49 | 856.02 | ---- | [43][Theo.] |
| | 9.66 | 902.03 | GGA(RPBE) | This work |
| | 9.77 | 932.57 | GGA(PBE) | [10][Theo.] |
| RbBi$_2$ | 9.65 | 898.63 | ---- | [17][Theo.] |
| | 9.61 | 887.50 | ---- | [43][Theo.] |

## 3.2. Elastic properties

Solids' elastic characteristics are closely related to atomic cohesion and cohesive energy. Elastic constants and moduli control how solids behave mechanically when they are under applied load. The elastic constants are the physical parameters used to measure a solid's mechanical strength and stability as well as to identify the type of interatomic bonding present in the substance. The elastic constants also have an impact on a material's mechanical characteristics, including stability, stiffness, brittleness, ductility, and elastic anisotropy. It is crucial to have knowledge of these characteristics when choosing a material for engineering reasons. Understanding a material's elastic characteristics is essential for technological development since it provides the knowledge necessary to understand how the material responds to various kinds of mechanical strains. Due to the great degree of symmetry in cubic systems, a cubic crystal's stiffness matrix includes only of the three distinct elastic constants: $C_{11}$, $C_{12}$, and $C_{44}$. The Born stability criteria [44] are inequality-based conditions that guarantee the mechanical stability of crystalline structures under static stress. Different crystal classes have different stability requirements. The Born stability condition, in the case of cubic crystals, simplify to the following form [45]:

$$C_{11} > 0, \; C_{44} > 0, \; (C_{11}-C_{12}) > 0, \; (C_{11}+2C_{12}) > 0 \qquad (8)$$

The computed values of $C_{ij}$ satisfy all the conditions stated above, ensuring that $X$Bi$_2$ ($X$ = K, Rb) compounds are elastically stable. According to our research, $C_{44}$, which stands for resistance to shear deformation, is less significant than $C_{11}$, which stands for unidirectional strain along the main crystallographic orientations. It is found that most of the elastic constants including $C_{44}$ are higher for RbBi$_2$ compared to those for KBi$_2$. Hardness of a solid depends on the value of $C_{44}$, therefore, it is predicted that RbBi$_2$ should be harder than KBi$_2$. The tetragonal shear modulus [46], $C'$, which is a measure of stiffness, is related to $C_{11}$ and $C_{12}$ by the following relation: $C' = (C_{11} - C_{12})/2$. This parameter is directly related to the crystal's dynamic stability and transverse acoustic wave velocity. Dynamical stability is suggested by a positive value of the tetragonal shear modulus of KBi$_2$ and RbBi$_2$.

The bulk modulus ($B$), shear modulus ($G$), Young's modulus ($Y$), and Poisson's ($\sigma$) ratio may all be determined using the Voigt-Reuss-Hill (VRH) approximation. The upper and lower limits of the elastic moduli of polycrystalline materials are determined via the Voigt and Reuss approximations, respectively. The real value is between the Voigt and Reuss limits. The bulk modulus ($B$) of a solid is a measure of its resistance to compression caused by constant hydrostatic pressure [47]. On the other side, the shear modulus (*G*) [48] demonstrates how effectively a solid can withstand external stress that tends to change the shape. The Young's modulus (*Y*) [49] measures how resistant a solid is to tensile stress that changes the length. The stiffness of a solid is also determined by this parameter. From the values of *Y* of the two compounds, it may be inferred that RbBi$_2$ is stiffer than KBi$_2$. The computed values are shown in Table 2. We can see from Table 2 that *B* > *G*, which suggests that the mechanical stability should be controlled by the shearing stress of *X*Bi$_2$ (*X* = K, Rb) compounds and that the mechanical failure mechanism of *X*Bi$_2$ (*X* = K, Rb) compounds will also be influenced more by the stress that changes shape than by the pressure affecting the volume.

In order to mimic the actual scenario, Hill [50] suggested utilizing the arithmetic mean of the two limits. The bulk modulus (*B*), shear modulus (*G*) (using the Voigt-Reuss-Hill (VRH) method), and Young's modulus (*Y*) were calculated using the following formulae [50–53]:

$$B_H = \frac{B_V + B_R}{2} \tag{9}$$

$$G_H = \frac{G_V + G_R}{2} \tag{10}$$

$$Y = \frac{9BG}{(3B + G)} \tag{11}$$

For most practical purposes, a solid is either brittle or ductile. Pugh's ratio (*B/G*), Poisson's ratio (*σ*), and Cauchy pressure (*C″*) are three important parameters that determine the brittle and ductile properties of a system [54,55]. Pugh proposed an empirical relationship to distinguish between the brittleness and ductility of materials. The Pugh's ratio, or the ratio of the bulk and shear moduli (*B/G*), determines the ductility or brittleness of a material. It is well known that a material will behave in a ductile manner if the Pugh's ratio value is more than 1.75, but in a brittle manner if the Pugh's ratio value is less than 1.75 [56]. In our case the (*B/G)* values are 4.85, and 1.85 for KBi$_2$, and RbBi$_2$ respectively, which suggests that the compounds should exhibit ductile behavior. The level of ductility is expected to be much higher for KBi$_2$.

The Poisson's ratio (*σ*) is a parameter used to describe how much a material has changed shape (either expanded or contracted) in relation to the direction in which it was loaded. The stability of solids against shear is also measured via this parameter. The possible values for Poisson's ratio of a solid is within the limit: -1.0 ≤ σ ≤ 0.5 [48]. If σ = 0.5, elastic deformation has no effect on volume. [57]. This parameter is capable of predicting whether a material will be brittle or ductile with a critical value of 0.26. If *σ* is less (greater) than 0.26, the material is brittle (ductile) [58–60]. From Table 2, we observe that the Poisson's ratios of KBi$_2$, and RbBi$_2$ are greater than 0.26. As a result, we can conclude that KBi$_2$, and RbBi$_2$ are ductile. This is in line with the outcome of the Pugh's ratio. Better plasticity results from a larger Poisson's ratio. The following expression expresses the Poisson's ratio:

$$\sigma = \frac{(3B - 2G)}{2(3B + G)} \tag{12}$$

By examining the value of σ [60,61], it is possible to study the nature of interatomic forces in solids. Central force interaction will predominate if this ratio stays between 0.25 and 0.50. If not, the non-central force will take control. Thus, in KBi$_2$, and RbBi$_2$, the bonding of atoms should be governed by a central force. For solids, the Cauchy pressure (*C″*) is another intriguing mechanical parameter. A material's Cauchy pressure is given by the formula *C″* = (*C$_{12}$* - *C$_{44}$*). Positive Cauchy pressure characterizes ductile materials, whereas negative Cauchy pressure characterizes brittle materials [61]. Cauchy pressure further clarifies the angular characteristics associated with atomic bonding in a material [62]. The existence of ionic and covalent bonding in a material is related to positive and negative values of the Cauchy pressure, respectively. According to the Pettifor's rule [62], a material with a significant positive Cauchy pressure and strong ductility has a lot of metallic bonding. Contrarily, a substance with negative Cauchy pressure has more angular bonds, which makes it more brittle and demonstrates considerable covalent bonding. As a result, positive value of Cauchy pressure indicates that $KBi_2$, and $RbBi_2$ are ductile in nature and have metallic bonds. Tetragonal shear modulus, Cauchy pressure, and independent elastic constants are given in Table 3.

There are several other important mechanical parameters of solids in view of their engineering applications, e.g., hardness (*H*) [63], machinability index ($\mu_M$) [64], and Kleinman parameter ($\zeta$) [65]. The concept known as "machinability" describes a property of a solid that affects how rapidly or readily it can be manufactured with a cutting tool. This characteristic is commonly used in engineering manufacture and production. The work material, the cutting tool, and the cutting parameters all have different effects on machinability [66]. The specific type of material utilized in the cutting instrument, as well as its form, cutting force, feed rate, and depth of cut, are all determined by the material's machinability. Additionally, it affects the flexibility and dry lubricating capabilities of the solids [67–69]. So, good machinability and improved dry lubricity are produced by combining strong bonding strength with low shear resistance. Materials with large value of $B/C_{44}$ feature high plastic strain values, low feed forces, good lubrication qualities, and low friction. Machinability index ($\mu_M$), has been calculated using the following equation:

$$\mu_M = \frac{B}{C_{44}} \qquad (13)$$

The Kleinman parameter ($\zeta$), also known as the internal strain parameter is a measurement that reveals how easily a material bends and stretches. Kleinman parameter ($\zeta$) has a value within 0 to 1. The lower value of $\zeta$ implies a considerable bond stretching or contracting contribution to resisting external stress, while the higher value reveals a significant bond bending reaction to resist external load [65]. From the value of Kleinman parameter ($\zeta$), it is evident that mechanical strength in $KBi_2$ and $RbBi_2$ is mainly derived from the bond stretching or contracting contribution. The following equation has been used to calculate the Kleinman parameter:

$$\zeta = \frac{C_{11} + 8C_{12}}{7C_{11} + C_{12}} \qquad (14)$$

Solids have a quality called hardness that describes how resistant they are to a localized plastic deformation. From the perspective of the application, the material's hardness is crucial. Understanding a material's mechanical and structural responses to severe stress requires knowledge of the material's hardness. In terms of elastic constants and moduli, $C_{44}$ and *G* are regarded as the best solids hardness indicators [70,71]. There are several expressions which can be used to compute the hardness of a solid. Some of the extensively used expressions are given in equations [(15) – (19)] below:

$$H_{micro} = \frac{(1-2\sigma)Y}{6(1+\sigma)} \qquad (15)$$

$$H_{macro} = 2\left[\left(\frac{G}{B}\right)^2 G\right]^{0.585} - 3 \qquad (16)$$

$$(H_V)_{Tian} = 0.92(G/B)^{1.137}G^{0.708} \qquad (17)$$

$$(H_V)_{Teter} = 0.151G \qquad (18)$$

$$(H_V)_{Mazhnik} = \gamma_0 \chi(\sigma) Y \qquad (19)$$

In Equation 5.17, $\chi(\sigma)$ is a function of Poisson's ratio and can be written as:

$$\chi(\sigma) = \frac{1 - 8.5\sigma + 19.5\sigma^2}{1 - 7.5\sigma + 12.2\sigma^2 + 19.6\sigma^3}$$

where, $\gamma_0$ is a dimensionless constant with a value of 0.096. The computed values of the hardness are displayed in Table 4.

**Table 2:** Calculated bulk modulus $B$ (in GPa), shear moduli $G$ (in GPa), Young modulus $Y$ (in GPa), Pugh's ratio ($B/G$), Poisson's ratio ($\sigma$), Cauchy pressure ($C''$) (in GPa), machinability index ($\mu_M$), Kleinman parameter ($\zeta$) for the $X$Bi$_2$ ($X$ = K, Rb) compounds.

| Compound | $B$ | $G$ | $Y$ | $B/G$ | $\sigma$ | $C''$ | $\mu_M$ | $\zeta$ | Ref. |
|---|---|---|---|---|---|---|---|---|---|
| KBi$_2$ | 26.86 | 5.54 | 15.54 | 4.85 | 0.41 | 17.42 | 4.07 | 0.89 | This work |
| RbBi$_2$ | 30.80 | 16.73 | 42.37 | 1.85 | 0.27 | 2.93 | 1.73 | 0.57 | This work |

**Table 3:** Calculated elastic constants, $C_{ij}$ (GPa), tetragonal shear modulus, $C'$ (GPa), and Cauchy pressure of the compounds $X$Bi$_2$ ($X$ = K, Rb).

| Compound | $C_{11}$ | $C_{12}$ | $C_{44}$ | ($C_{11}$ - $C_{12}$) | $C'$ | Ref. |
|---|---|---|---|---|---|---|
| KBi$_2$ | 32.54 | 24.02 | 6.60 | 8.52 | 4.26 | This work |
| RbBi$_2$ | 50.93 | 20.74 | 17.81 | 30.19 | 15.09 | This Work |

**Table 4:** Calculated hardness (GPa) based on elastic moduli and Poisson's ratio of $X$Bi$_2$ ($X$ = K, Rb) compounds.

| Compound | $H_{micro}$ | $H_{macro}$ | $(H_V)_{Tian}$ | $(H_V)_{Teter}$ | $(H_V)_{Mazhnik}$ |
|---|---|---|---|---|---|
| KBi$_2$ | 0.33 | -0.84 | 0.51 | 0.84 | 0.89 |
| RbBi$_2$ | 2.57 | 2.08 | 3.38 | 2.53 | 2.08 |

It is found from Table 3 that KBi$_2$ is very highly machinable. Both KBi$_2$ and RbBi$_2$ are very soft in nature and all the elastic constants and moduli of these two Laves phase compounds are low.

3.3. Elastic anisotropy

Elastic anisotropy affects all the mechanical phenomena, including the growth of plastic deformations in crystals, the movement of fractures in materials, and the creation of cracks in materials. An anisotropic material is one whose physical characteristics differ in different directions. Understanding elastic anisotropy has significant ramifications for both applied engineering sciences and crystal physics. As a result, calculation of the elastic anisotropy parameters is essential for $X$Bi$_2$ ($X$ = K, Rb) compounds in detail to learn more about its adaptability and possible applications under numerous circumstances of external stress.

In our investigations, we have identified the Zener anisotropy factor ($A$), shear anisotropy factors ($A_1$, $A_2$, $A_3$), and percentage anisotropy in compressibility ($A_B$) and shear ($A_G$). The results of the calculation are revealed in Table 5. Additionally, we have estimated the universal anisotropy factor ($A^U$, $d_E$), equivalent Zener anisotropy factor ($A^{eq}$), the universal log-Euclidean index ($A^L$), the anisotropies of the bulk modulus along $a$-, $b$- and $c$-axis ($A_{B_a}$ and $A_{B_c}$) and they are also given in Table 5. To determine these anisotropy factors, the following formulas are applied [36-41,72]:

$$B_a = a\frac{dP}{da} = \frac{\Lambda}{1+\alpha+\beta} \tag{20}$$

$$B_b = a\frac{dP}{db} = \frac{B_a}{\alpha} \tag{21}$$

$$B_c = c\frac{dP}{dc} = \frac{B_a}{\beta} \tag{22}$$

$$A_{B_a} = \frac{B_a}{B_b} = \alpha \tag{23}$$

$$A_{B_c} = \frac{B_c}{B_b} = \frac{\alpha}{\beta} \tag{24}$$

where, $\Lambda = C_{11} + 2C_{12}\alpha + C_{22}\alpha^2 + 2C_{13}\beta + C_{33}\beta^2 + 2C_{33}\alpha\beta$ and regarding cubic crystals, $\alpha = \beta = 1$. $A_{B_a} = A_{B_c} = 1$, indicates isotropy in the bulk modulus. Consequently, the directional bulk modulus is isotropic in KBi$_2$, and RbBi$_2$ (Table 5). Estimated directional bulk moduli for various crystallographic axes of KBi$_2$, and RbBi$_2$ compounds are provided equally in Table 5. These numbers are larger than the bulk modulus of isotropic polycrystalline crystals. This results from the realization that for a specific crystal density, the amount of pressure in a state of uniaxial strain typically is different from the pressure in a condition of hydrostatic stress at the same density of the solid [70].

Three variables may be used to calculate the cubic crystal's shear anisotropy factor [72,73]:

Shear anisotropy for {100} shear planes in the ⟨011⟩ and ⟨010⟩ directions, $A_1$:

$$A_1 = \frac{4C_{44}}{C_{11}+C_{33}-2C_{13}} \tag{25}$$

Shear anisotropy for {010} shear planes in the ⟨101⟩ and ⟨001⟩ directions, $A_2$:

$$A_2 = \frac{4C_{55}}{C_{22}+C_{33}-2C_{23}} \tag{26}$$

Shear anisotropy for {001} shear planes in the ⟨110⟩ and ⟨010⟩ directions, $A_3$:

$$A_3 = \frac{4C_{66}}{C_{11}+C_{22}-2C_{12}} \tag{27}$$

The computed values of these anisotropy factors are enlisted in Table 5. The computed values of $A_1$, $A_2$, and $A_3$ are significantly different from 1. The percentage anisotropy in the bulk modulus is very low for $KBi_2$, and $RbBi_2$ because of the cubic structure. From Table 5, it can be observed that the anisotropic behavior of $KBi_2$ is larger than that of $RbBi_2$. Zener anisotropy factor, $A$, can be obtained from the expression given below [74]:

$$A = \frac{2C_{44}}{C_{11}-C_{12}} \tag{28}$$

The log-Euclidean formula is used to determine the universal log-Euclidean index [74,75]:

$$A^L = \sqrt{[\ln(\frac{B_V}{B_R})]^2 + 5[\ln(\frac{C_{44}^V}{C_{44}^R})]^2} \tag{29}$$

The Voigt and Reuss values of $C_{44}$ are obtained in this approach from [76]:

$$C_{44}^V = C_{44}^R + \frac{3}{5}\frac{(C_{11}-C_{12}-2C_{44})^2}{3(C_{11}-C_{12})+4C_{44}} \tag{30}$$

$$C_{44}^R = \frac{5}{3}\frac{C_{44}(C_{11}-C_{12})}{3(C_{11}-C_{12})+4C_{44}} \tag{31}$$

Kube and Jong [74,75] stated that for inorganic crystalline compounds, $0 \leq A^L \leq 10.26$, and 90% of these compounds have $A^L < 1$. $A^L = 0$ suggests complete isotropy. Moreover, most (78%) of these inorganic crystalline compounds with high $A^L$ values have layered or lamellar structures [76]. Higher $A^L$ values indicate strongly layered structural elements, whereas lower $A^L$ values indicate a non-layered structure. Considering the substantially lower value of $A^L$, we can anticipate that $KBi_2$, and $RbBi_2$ display a layout that is moderately layered. To determine the universal anisotropy index, $(A^U, d_E)$, equivalent Zener anisotropy measure, $A^{eq}$, percentage anisotropy in compressibility, $A_B$ and anisotropy in shear, $A_G$ (or $A_C$) of the compounds of interest, the following common equations are used [74,77–79]:

$$A^U = 5\frac{G_V}{G_R} + \frac{B_V}{B_R} - 6 \geq 0 \tag{32}$$

$$d_E = \sqrt{A^U + 6} \tag{33}$$

$$A^{eq} = \left(1 + \frac{5}{12}A^U\right) + \sqrt{(1+\frac{5}{12}A^U)^2 - 1} \tag{34}$$

$$A_B = \frac{B_V - B_R}{B_V + B_R} \tag{35}$$

$$A_G = \frac{G_V - G_R}{2G_H} \tag{36}$$

The parameter $A^U$ is called the universal anisotropy index, because of it applies to a variety of crystal symmetries. The requirement for an isotropic crystal is $A^U = 0$. Anisotropy can be determined from this value, which needs to be positive. $A^U$ for KBi$_2$, and RbBi$_2$ displays the same trend as $A^L$. For $A^{eq}$, any other value than unity implies anisotropy. The determined values of $A^{eq}$ for KBi$_2$, and RbBi$_2$ compounds that are listed in Table 5, anticipate that KBi$_2$, and RbBi$_2$ are moderately anisotropic. For $A_B$ and $A_G$, elastic anisotropy has a value of 1, while a value of 0 corresponds to elastic isotropy [80]. It is noteworthy that $A_G$ predicts a significantly lower value for the anisotropy index than do other measurements. The cubic structure of KBi$_2$ and RbBi$_2$ compounds and a zero value of $A_B$ for these compounds demonstrate that the bulk modulus has no impact on the anisotropic elastic and mechanical properties. Additionally, it is shown that KBi$_2$ exhibits somewhat more anisotropic elastic and mechanical behavior than RbBi$_2$.

**Table 5:** Shear anisotropy factors ($A_1$, $A_2$ and $A_3$), Zener anisotropy factor $A$, universal log-Euclidean index $A^L$, the universal anisotropy index ($A^U$, $d_E$), equivalent Zener anisotropy measure $A^{eq}$, anisotropy in shear $A_G$, anisotropy in compressibility $A_B$, anisotropy in bulk modulus, and directional bulk modulus (GPa) for $X$Bi$_2$ ($X$ = K, Rb) compounds.

| Compound | $A_1$ | $A_2$ | $A_3$ | $A$ | $A^L$ | $A^U$ | $d_E$ | $A^{eq}$ | $A_G$ | $A_B$ | $A_{B_a}$ | $A_{B_c}$ | $B_a$ |
|---|---|---|---|---|---|---|---|---|---|---|---|---|---|
| KBi$_2$ | 1.55 | 1.55 | 1.55 | 1.55 | 0.74 | 0.23 | 2.50 | 1.54 | 0.02 | 0 | 1 | 1 | 86.26 |
| RbBi$_2$ | 1.18 | 1.18 | 1.18 | 1.18 | 0.44 | 0.033 | 2.46 | 1.18 | 0.003 | 0 | 1 | 1 | 112.54 |

Using the ELATE code, we have shown the variation of Young's modulus ($Y$), compressibility ($\beta$), shear modulus ($G$), and Poisson's ratio ($\sigma$) which further illustrates elastic anisotropy. For an isotropic crystal, the 3D plots must be spherical; otherwise, they show the degree of anisotropy. We have shown ELATE [81] generated 2D and 3D plots of directional dependences of the Young's modulus, shear modulus, linear compressibility, and Poisson's ratio for $X$Bi$_2$ ($X$ = K, Rb) compounds below. From Figure 2 to Figure 5, it is clear that the 3D figures of $Y$, $G$, and $\sigma$ show a slight departure from spherical shape, indicating some degree of anisotropy. However, the 3D figure of $\beta$ representing the isotropy as it does not deviate from the spherical shape. The ELATE-generated graphs clearly reveal that the direction-dependent $Y$, $\beta$, $G$, and $\sigma$ projections in the *ab*-plane are nearly circular. This suggests that the basal plane's elastic anisotropy is comparatively insignificant. Table 6 lists the maximum and minimum values of $Y$, $\beta$, $G$, and $\sigma$ as well as their maximum to minimum ratios. These proportions provide helpful approximations for elastic anisotropy. According to Table 6, the order of anisotropy in the elastic properties of KBi$_2$, and RbBi$_2$ compounds are listed in the following order: KBi$_2$ > RbBi$_2$. Other anisotropy measurements in the earlier parts also produce similar outcomes.

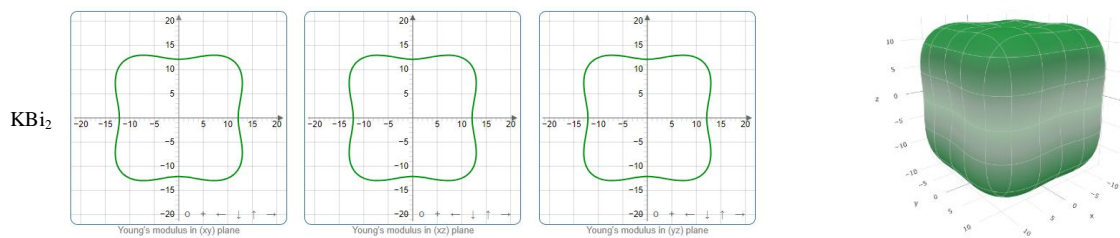

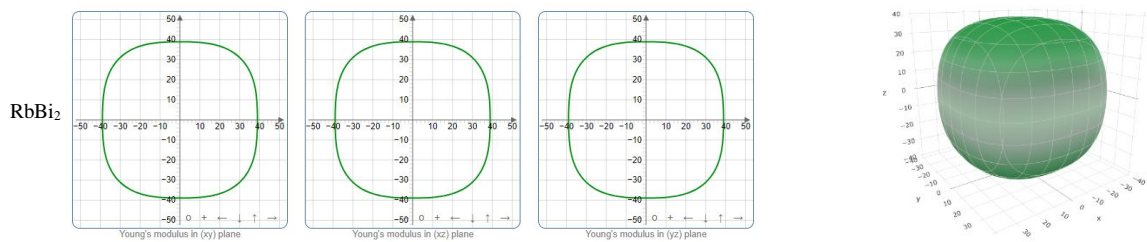

**Figure 2:** 2D and 3D directional dependences in Young's modulus (*Y*) of $X$Bi$_2$ ($X$ = K, Rb) compounds.

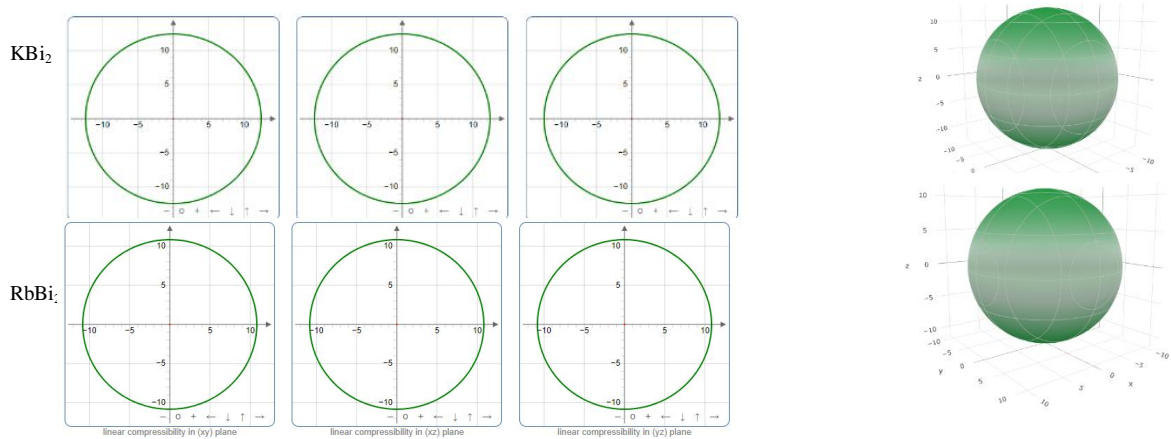

**Figure 3:** 2D and 3D directional dependences in linear compressibility (*β*) of $X$Bi$_2$ ($X$ = K, Rb) compounds.

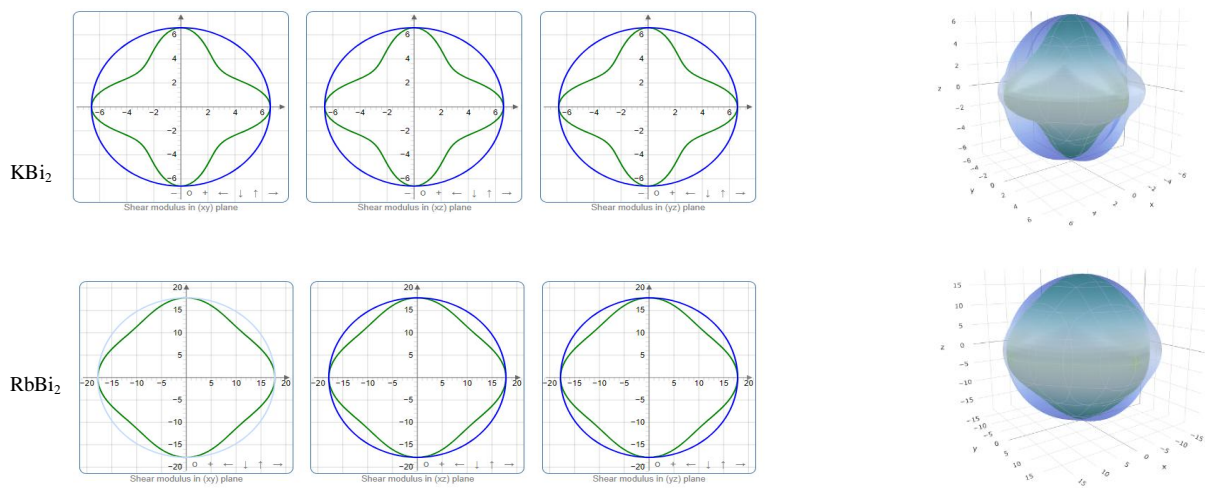

**Figure 4:** 2D and 3D directional dependences in shear modulus (*G*) of $X$Bi$_2$ ($X$ = K, Rb) compounds.

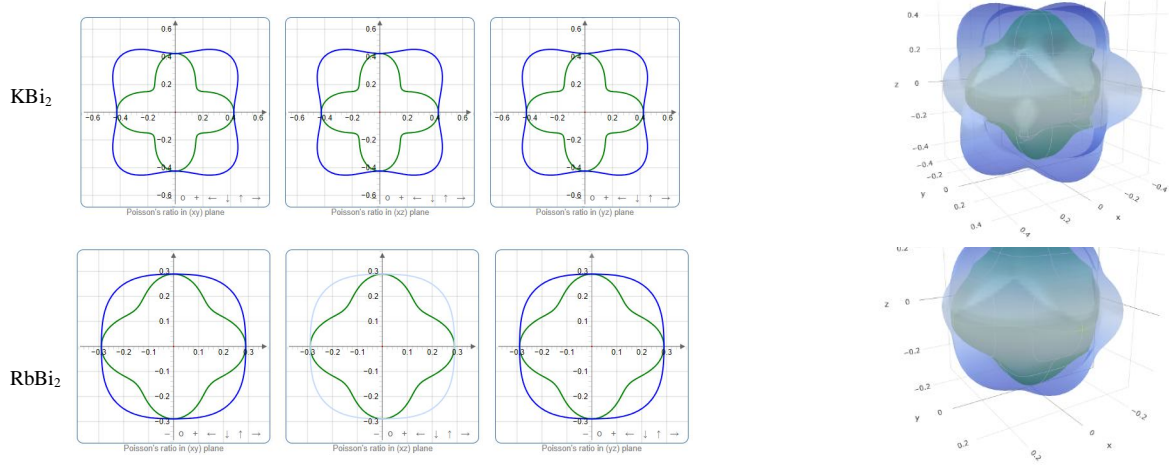

**Figure 5:** 2D and 3D directional dependences in Poisson's ratio ($\sigma$) of $X$Bi$_2$ ($X$ = K, Rb) compounds.

**Table 6:** The maximum and minimum values of Young's modulus $Y$ (GPa), compressibility $\beta$ (TPa$^{-1}$), shear modulus $G$ (GPa), Poisson's ratio $\sigma$, and their ratios for $X$Bi$_2$ ($X$ = K, Rb) compounds.

| Compound | Y | | $A_Y$ | $\beta$ | | $A_\beta$ | G | | $A_G$ | $\sigma$ | | $A_\sigma$ |
|---|---|---|---|---|---|---|---|---|---|---|---|---|
| | $Y_{min}$ | $Y_{max}$ | | $\beta_{min}$ | $\beta_{max}$ | | $G_{min}$ | $G_{max}$ | | $\sigma_{min}$ | $\sigma_{max}$ | |
| KBi$_2$ | 12.13 | 18.30 | 1.51 | 12.41 | 12.41 | 1.00 | 4.26 | 6.60 | 1.55 | 0.23 | 0.57 | 2.48 |
| RbBi$_2$ | 38.93 | 44.81 | 1.15 | 10.82 | 10.82 | 1.00 | 15.09 | 17.82 | 1.18 | 0.21 | 0.32 | 1.52 |

### 3.4. Electronic band structure and energy density of states

(a) Band structure

The band structure of a solid can be used to describe its optical and charge transport characteristics. The band structure also contains information regarding bonding and electronic stability. We have calculated the electronic band structure for the $X$Bi$_2$ ($X$ = K, Rb) compounds in the k-space along high symmetry directions as a function of energy $E$ (-5 eV to 10 eV). The band dispersions are displayed in Figure 6. The horizontal broken line represents the Fermi level $E_F$ in Figure 6. The band structure indicates the metallic nature of KBi$_2$, and RbBi$_2$ compounds as at the Fermi level, conduction bands and valance bands are clearly overlapping. Strong band overlap and Fermi level crossings are present. The number of bands crossing the Fermi level is 10 for both the compounds. Highly dispersive bands are crossing the Fermi level near $\Gamma$-point and $L$-point for $X$Bi$_2$ ($X$ = K, Rb) compounds. Highly dispersed bands are indicative of significant charge carrier mobility in the corresponding region of the Brillouin zone. Low effective mass of charge carriers and strong mobility are suggested by highly dispersive bands [82–84]. The qualitative features of the band structures of $X$Bi$_2$ ($X$ = K, Rb) compounds are quite similar close to the Fermi level (Fig. 6).

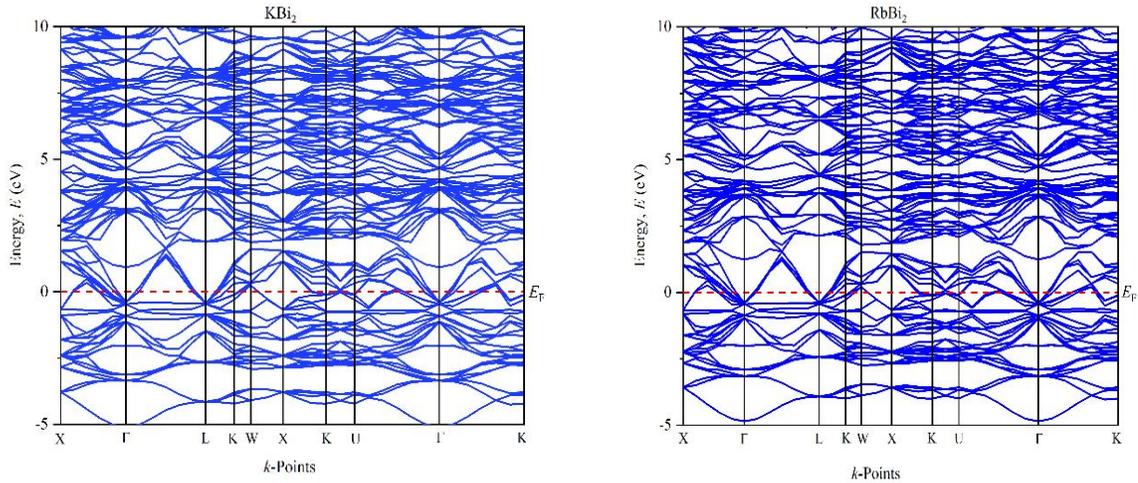

**Figure 6:** The electronic band structure of $X\text{Bi}_2$ ($X$ = K, Rb) compounds in the first Brillouin zone.

(b) Electronic energy density of states (DOS)

The total number of possible energy states that electrons can occupy per unit of energy interval is described by the electronic density of states (EDOS). The DOS is essential to solid state physics because it controls numerous electrical, optical, and bonding properties. A high DOS at a given energy level means that there are many states that are available for occupation. A DOS of zero means that at that particular energy, no electronic states are accessible. Total density of states (TDOS) and partial density of states (PDOS) of $X\text{Bi}_2$ ($X$ = K, Rb) compounds have been calculated to further explain the significance of various orbitals to the electronic characteristics and nature of chemical bonding (displayed in Figure 7). The vertical dashed line shows the Fermi level, $E_F$, at 0 eV for $\text{KBi}_2$ and $\text{RbBi}_2$ compounds. The metallic nature of $\text{KBi}_2$ and $\text{RbBi}_2$ compounds can be seen by the non-zero TDOS at the Fermi level, that was entirely consistent with the features of the electronic band structure. At the Fermi level, the TDOS for $\text{KBi}_2$ is 6.28 electronic states per eV and for $\text{RbBi}_2$ it is 5.21 electronic states per eV. The Bi-6p orbitals for $\text{KBi}_2$ and $\text{RbBi}_2$ compounds make up the majority of the near-Fermi level contribution to the TDOS. Thus, the electrical conductivity of the compounds $\text{KBi}_2$ and $\text{RbBi}_2$ should be dominated by this particular electronic state. The Bi-6p electrons should also play a significant role in the structural and chemical stabilities of the molecules $\text{KBi}_2$ and $\text{RbBi}_2$. Previous studies found that both metallic and superconducting properties in $\text{KBi}_2$, and $\text{RbBi}_2$ compounds arise from the 6p states of Bi atoms [17,85]. The peak closest to the negative energy below the Fermi level in TDOS is called the bonding peak, while the peak closest to the positive energy is called the anti-bonding peak. Electrical stability is indicated by the pseudogap, which is the energy difference between these peaks [86–88]. The TDOS of $\text{KBi}_2$, and $\text{RbBi}_2$ compounds show the $E_F$ is close to the anti-bonding peak. In $\text{KBi}_2$, and $\text{RbBi}_2$ compounds bonding and anti-bonding peaks are within 1.14 eV, and 1.07 eV, respectively, from the Fermi level. A material's structural stability is greatly influenced by the interaction of charges between bonding atoms; materials with more bonding electrons have more stable structures [89,90].

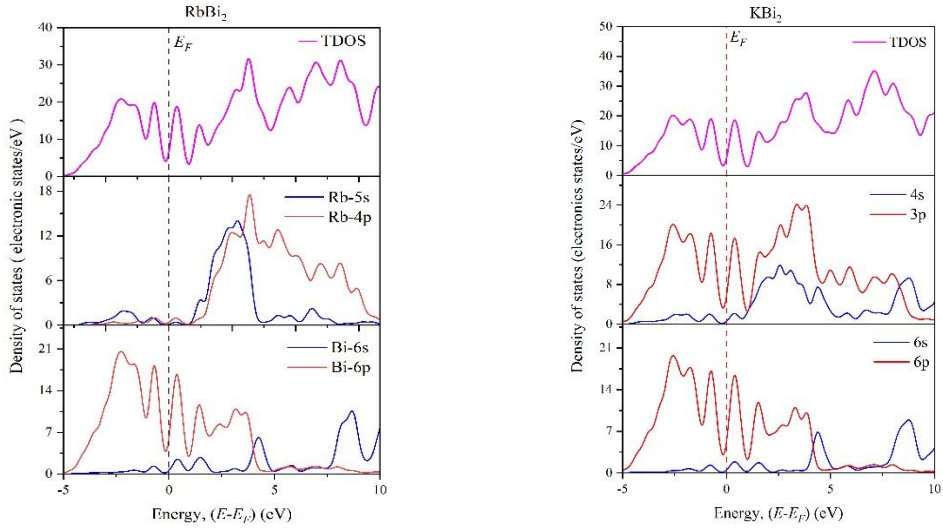

**Figure 7:** Total and partial density of states of $X$Bi$_2$ ($X$ = K, Rb) compounds.

## 3.5. Optical properties

The frequency or energy dependent optical properties are important for the understanding of the electronic energy band structure, impurity level states, excitons, and localized defect states [91,92]. Additionally, thorough knowledge of a system's sensitivity to infrared, visible, and ultraviolet spectra is helpful in order to find potential optoelectronic applications [93,94]. We have used 0.5 eV Gaussian smearing in all the optical calculations, an empirical Drude term with a plasma frequency of 8 eV and damping of 0.05 eV were also.

In this section, the frequency/energy dependent optical constants, namely, the complex dielectric function ε(ω), refractive index n(ω), optical conductivity σ(ω), reflectivity R(ω), absorption coefficient α(ω), and energy loss function L(ω), are disclosed and discussed. Figure 8, and Figure 9 show the optical characteristics of the compounds of interest for electric polarizations along the [100] direction and for incident energy up to 20 eV.

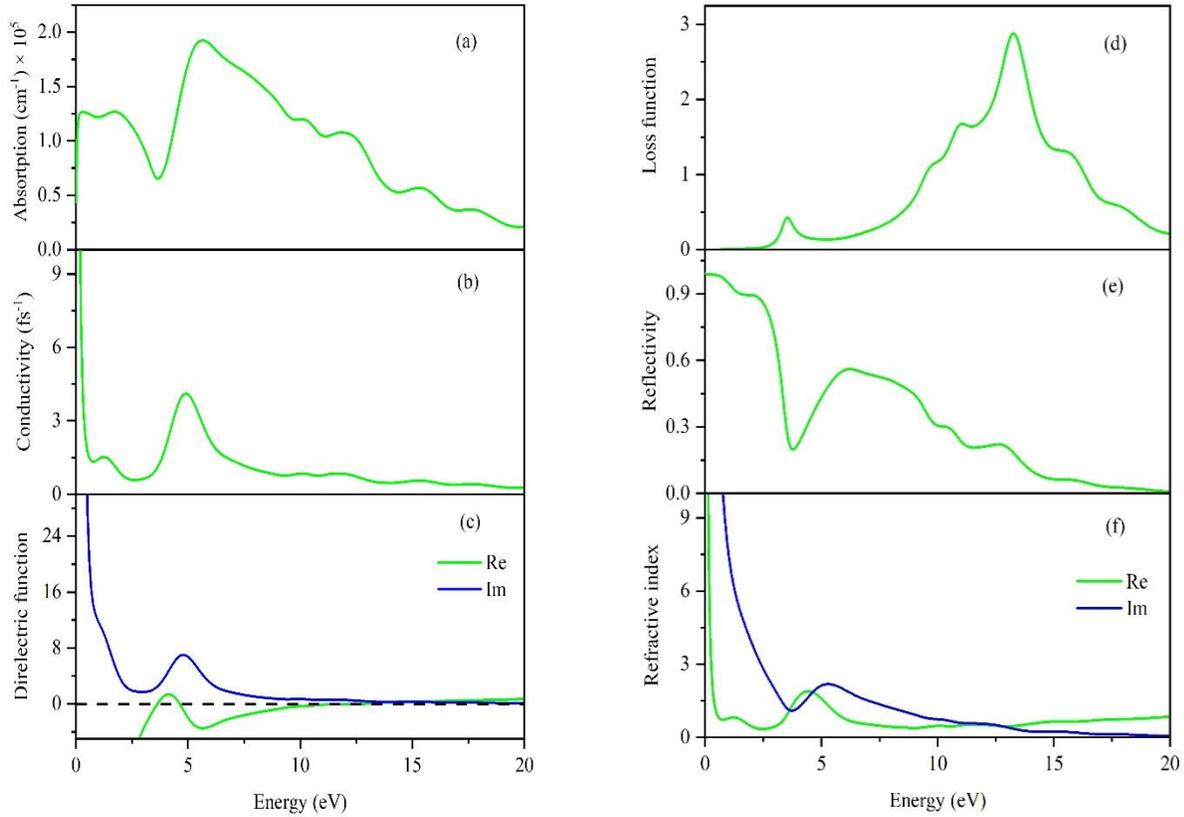

**Figure 8:** The energy-dependent (a) absorption coefficient (b) optical conductivity (c) dielectric function (d) loss function (e) reflectivity, and (f) refractive index of $KBi_2$.

The absorption coefficient $\alpha(\omega)$ determines how far light of a specific energy (wavelength/frequency) can penetrate into the material before being absorbed. The absorption coefficients $\alpha(\omega)$ of the of $KBi_2$, and $RbBi_2$ compounds are shown in Figure 8(a) and 9(a). The absorption coefficients of $KBi_2$ and $RbBi_2$ compounds start at 0 eV, indicating that it has a metallic band structure. $KBi_2$ has a rather high absorption coefficient in the spectrum range from ~0 to 12.05 eV, with a maximum value around ~5.67 eV in the ultraviolet region. On the other hand, $RbBi_2$ has a quite high absorption coefficient in the spectral range of ~0 to 11.81 eV, with a highest value at approximately ~12.7 eV also in the ultraviolet region. Absorption coefficient $\alpha(\omega)$ decreases from ~12.61 eV, and ~12.22 eV for $KBi_2$, and $RbBi_2$, respectively.

The optical conductivity of charge carriers characterizes their dynamic response to incident light. The real and imaginary parts of optical conductivity $\sigma(\omega)$ are depicted in Figure 8(b), and 9(b). The optical conductivity of $KBi_2$ and $RbBi_2$ compounds starts at zero photon energy, demonstrating the absence of a band gap, which is consistent with the electronic band structure and TDOS calculations and supports the metallic nature of the materials under investigation.

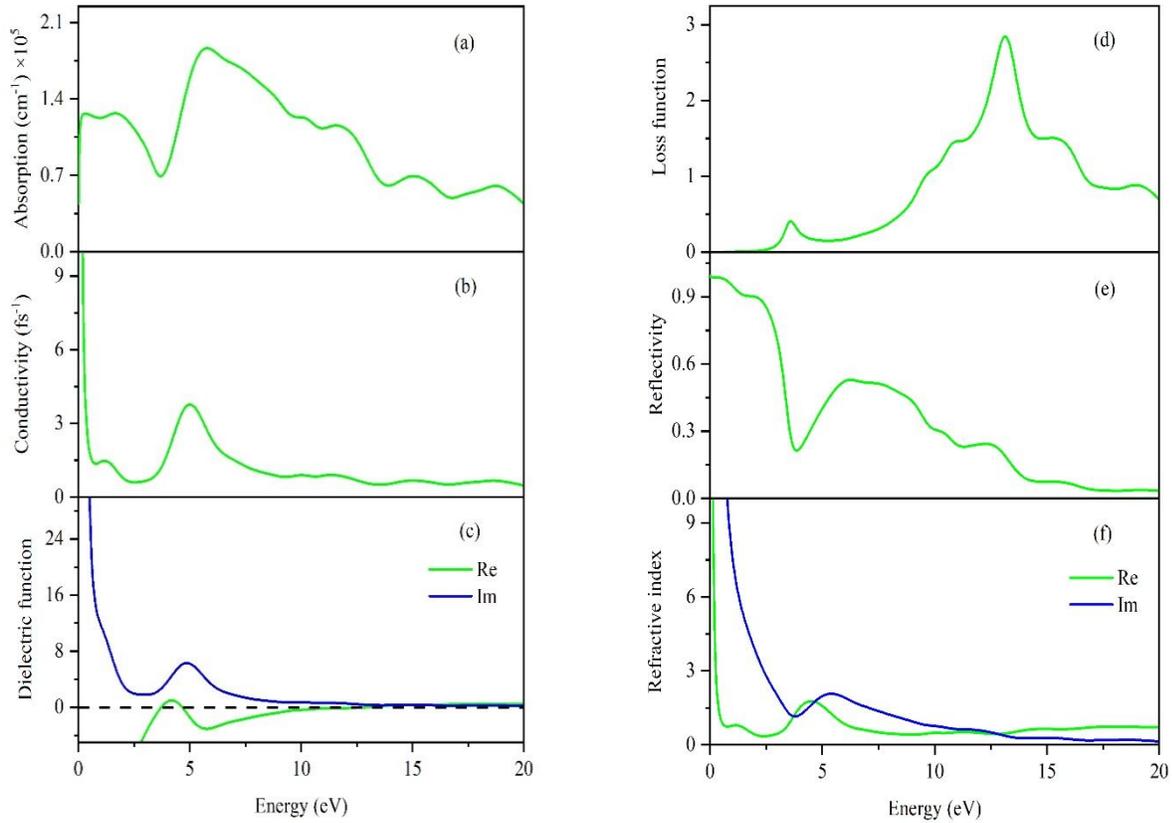

**Figure 9:** The energy-dependent (a) absorption coefficient (b) optical conductivity (c) dielectric function (d) loss function (e) reflectivity, and (f) refractive index of $RbBi_2$.

Figure 8(c) and 9(c) illustrate the real and imaginary components of the dielectric constants of $KBi_2$ and $RbBi_2$ compounds. The imaginary part of the dielectric function has a connection to the dissipation of energy of the electromagnetic wave within the medium, whereas the real part is associated with electrical polarization and anomalous dispersion. At ~3.63 eV for $KBi_2$ and ~3.71 eV for $RbBi_2$, the real part of the dielectric function crosses zero from below. This ensures metallic feature of the compounds. The energy at which the dielectric loss inside a particular compound reaches its maximum is indicated by the peak in the imaginary component. The magnitudes of both real and imaginary components decrease monotonically in the high energy region.

Figure 8(d), and 9(d) show the loss functions $L(\omega)$ of $KBi_2$, and $RbBi_2$ compounds. The loss function $L(\omega)$ describes the energy loss of a fast electron via exciting collective charge oscillation modes while traversing a material [95,96]. The loss peaks for $KBi_2$ are found at ~13.26 eV, and $RbBi_2$ at ~13.15 eV. The positions of the loss peaks represent the plasmon energy. At this particular energy, plasma oscillations are induced due to collective oscillation of electrons in the jellium model. It is significant to note that the plasma energies are accompanied by sudden reductions in the reflectance and absorption

coefficient. Both the Laves phase compounds become transparent to incident photons above the plasma energy.

The amount of energy an electromagnetic wave (EMW) reflected from a surface compared to the energy it had when it first hit the surface is known as the reflectivity. Figure 8(e) and 9(e) illustrate the reflectivity profiles for KBi$_2$, and RbBi$_2$ compounds. At 0 eV, the maximum reflectivity value in KBi$_2$ is around ~98%. From infrared to near-ultraviolet (up to ~2.21 eV), the reflectivity spectra stay over ~89%. At 0 eV, the maximum reflectivity value in RbBi$_2$ is around ~99%. From infrared to near-ultraviolet (up to ~2.11 eV), the reflectivity spectra stay over ~90%. Such extremely high reflectivity is due to the highly metallic nature of the compounds under study. The reflectivity spectra of the binary Laves phases indicate that these two compounds are extremely efficient solar energy reflector.

Figure 8(f) and 9(f) show the real and imaginary parts of the refractive index of KBi$_2$ and RbBi$_2$. The real part of the refractive index governs the phase velocity of an EMW, whereas the imaginary part, also known as the extinction coefficient, governs the attenuation of the EMW within the material. It is significant that at low energies both KBi$_2$ and RbBi$_2$ exhibit high values of the real component of the refractive index which is in the infrared regions. Such high values are useful in optoelectronic device applications [41].

The optical parameters presented in this section are completely novel. In terms of the polarization state of the incident electric field, all of the optical spectra exhibit a significant degree of isotropies because of the cubic structure of KBi$_2$ and RbBi$_2$.

### 3.6. Thermo-physical properties

(a) Debye temperature

Debye temperature ($\Theta_D$) is important for understanding a variety of thermo-physical properties of solids, such as their bonding forces, energy of vacancy formation, melting point, thermal conductivity, phonon dynamics, specific heat, and superconductivity, among others. The Debye temperature is commonly thought of as the temperature at which a solid's phonon wavelengths approximately correspond to that of its interatomic distance. This temperature is used to split the lattice dynamics of a solid into high- and low-temperature zones. The classical and quantum behaviors of the lattice vibration are also separated by the Debye temperature. All vibrational modes have the same energy, ~k$_B$T, when T > $\Theta_D$. However, the higher frequency modes are absent at T < $\Theta_D$ [97]. There are various approaches of estimating $\Theta_D$. Low temperature vibrational excitations brought on by acoustic modes produce agreement between specific heat measurements and the Debye temperature calculated from the elastic constants. Here, we have used the following formalism to calculate the Debye temperature of XBi$_2$ (X = K, Rb) [97]:

$$\Theta_D = \frac{h}{k_B}\left[\left(\frac{3n}{4\pi}\right)\frac{N_A\rho}{M}\right]^{\frac{1}{3}} v_m \tag{37}$$

where $h$ is the Planck's constant, $k_B$ is the Boltzmann's constant, $n$ denotes the number of atoms within the unit cell, $M$ is molar mass, $\rho$ is density, $N_A$ is Avogadro's number and $v_m$ denotes mean sound velocity. $v_m$ can be determined from bulk and shear moduli through the longitudinal ($v_l$) and transverse ($v_t$) sound velocities as follows:

$$v_m = \left[\frac{1}{3}\left(\frac{2}{v_t^3} + \frac{1}{v_l^3}\right)\right]^{-\frac{1}{3}} \tag{38}$$

where,

$$v_t = \sqrt{\frac{G}{\rho}} \tag{39}$$

$$v_l = \sqrt{\frac{3B+4G}{3\rho}} \tag{40}$$

It is known that a higher phonon thermal conductivity corresponds to a higher Debye temperature. Higher Debye temperatures are encouraged by strong chemical bonds. As shown in Table 7, in comparison to KBi$_2$, we find that RbBi$_2$ has a greater $\Theta_D$ value. The estimated $\Theta_D$ of $X$Bi$_2$ ($X$ = K, Rb) compounds are very low. This implies weak interatomic bonding between the atoms involved and overall soft nature of the compounds.

**Table 7:** Calculated mass density ($\rho$ in gm cm$^{-3}$), longitudinal, transverse, average sound velocities ($v_l$, $v_t$ and $v_m$ in ms$^{-1}$), and Debye temperature ($\Theta_D$ in K) of $X$Bi$_2$ ($X$ = K, Rb) compounds.

| Compound | $\rho$ | $v_l$ | $v_t$ | $v_m$ | $\Theta_D$ | Ref. |
|---|---|---|---|---|---|---|
| KBi$_2$ | 7.02 | 2210.12 | 890.01 | 1008.71 | 91.05 | This work |
| RbBi$_2$ | 7.41 | 2671.59 | 1497.18 | 1659.41 | 147.65 | This work |

(b) Sound velocities

The elastic characteristics have a substantial impact on how elastic waves propagate through a material. The material's acoustic characteristics also significantly affect electrical and thermal conductivities via the phonon dynamics. Today, research into the properties of elastic waves is crucial in the fields of geology, materials science, seismology, and medicine. Each atom in a solid only vibrates in one longitudinal and two transverse modes, totaling just three modes. In crystals, there are just a few crystallographic directions where the pure longitudinal and transverse modes are present. To understand the anisotropic nature of sound velocity, measurements in several propagation directions should be made. Pure transverse and longitudinal modes can be identified for [001], [110], and [111] directions with cubic symmetry; sound propagation modes in other directions are quasi-transverse or quasi-longitudinal. The acoustic velocities in the principle directions for cubic crystals can be expressed as [80]:

$$[100]v_l = \sqrt{C_{11}/\rho}; \quad [010]v_{t1} = [001]v_{t2} = \sqrt{C_{44}/\rho} \tag{41}$$

$$[110]v_l = \sqrt{(C_{11} + C_{12} + 2C_{44})/2\rho}; \quad [1\bar{1}0]v_{t1} = \sqrt{(C_{11} - C_{12})/\rho}; \quad [001]v_{t2} = \sqrt{C_{44}/\rho} \tag{42}$$

$$[111]\upsilon_l = \sqrt{(C_{11} + 2C_{12} + 4C_{44})/3\rho}; [11\bar{2}]\upsilon_{t1} = \upsilon_{t2} = \sqrt{[(C_{11} - C_{12}) + C_{44}]/3\rho} \qquad (43)$$

The terms "$\upsilon_{t1}$" and "$\upsilon_{t2}$" denote, respectively, the first and second transverse modes. A compound that has a low density $\rho$ and high elastic constants has high sound velocities. So, we can say that RbBi$_2$ has larger sound velocities than KBi$_2$. Directional sound velocities of $X$Bi$_2$ ($X$ = K, Rb) compounds are disclosed in Table 8.

**Table 8:** Anisotropic sound velocities (ms$^{-1}$) of $X$Bi$_2$ ($X$ = K, Rb) compounds along different crystallographic directions.

| Propagation directions and modes | | KBi$_2$ | RbBi$_2$ |
|---|---|---|---|
| [111] | $[111]\upsilon_l$ | 2252.23 | 2659.93 |
| | $[11\bar{2}]\upsilon_{t1}$ | 926.92 | 1469.44 |
| | $[11\bar{2}]\upsilon_{t2}$ | 926.92 | 1469.44 |
| [110] | $[110]\upsilon_l$ | 2227.46 | 2690.64 |
| | $[1\bar{1}0]\upsilon_{t1}$ | 1100.89 | 2018.47 |
| | $[001]\upsilon_{t2}$ | 968.93 | 1550.32 |
| [100] | $[100]\upsilon_l$ | 2151.45 | 2621.67 |
| | $[010]\upsilon_{t1}$ | 968.93 | 1550.32 |
| | $[001]\upsilon_{t2}$ | 968.93 | 1550.32 |

Sound velocities in RbBi$_2$ for all the modes are significantly higher compared to those for KBi$_2$. This is due to relatively higher level of stiffness of RbBi$_2$.

(c) Lattice thermal conductivity

Thermal energy can be transported by both phonons and electrons in materials. At low temperatures, electrons are the main heat transporters in metals. The lattice plays a more important part at high temperatures. Investigating a material's lattice thermal conductivity ($k_{ph}$) is crucial for high-temperature applications. The amount of heat energy transported in the presence of temperature gradients via lattice vibration depends on a solid's lattice thermal conductivity ($k_{ph}$). Both low and high lattice thermal conductivity materials are used in several engineering specialties. In an effort to improve the performance of solid state refrigeration, thermoelectric devices, and thermal barrier coatings (TBC), the quest for low thermal conducting materials has recently attracted attention. On the other hand, to increase the effectiveness of heat removal in microelectronic and nanoelectronic devices, excellent thermal conductivity materials with a small amount of heat waste are required. There is a formula that can be used to estimate the $k_{ph}$ as a function of temperature derived by Slack [98] as follows:

$$k_{ph}(T) = A \frac{M_{av}\Theta_D^3 \delta}{\gamma^2 n^{2/3} T} \qquad (44)$$

In this formulation, $M_{av}$ is the average atomic mass in kg/mol, $\Theta_D$ is the Debye temperature in K, $\delta$ is the cubic root of average atomic volume in meter, $n$ refers to the number of atoms in the conventional unit cell, $T$ is the absolute temperature in K, and $\gamma$ is the acoustic Grüneisen parameter ($\gamma$) which determines the degree of anharmonicity of phonons. Grüneisen parameter ($\gamma$) is a crucial parameter in lattice dynamics and thermodynamics due to its relationships with the volume, bulk modulus, heat capacity, and thermal expansion coefficient. Low phonon anharmonicity, which results in good thermal conductivity, is a property of materials having low Grüneisen parameter ($\gamma$) values. It means that RbBi$_2$ has better thermal conductivity than KBi$_2$ as its Grüneisen parameter is lower. It is a dimensionless quantity which can be derived from the Poisson's ratio using the equation [99]:

$$\gamma = \frac{3(1+v)}{2(2-3v)} \qquad (45)$$

The factor $A(\gamma)$ due to Julian [99] is calculated from:

$$A(\gamma) = \frac{5.720 \times 10^7 \times 0.849}{2 \times (1 - 0.514/\gamma + 0.228/\gamma^2)} \qquad (46)$$

The calculated room temperature (300 K) lattice thermal conductivity is listed in Table 9. The lattice thermal conductivity ($k_{ph}$) of RbBi$_2$ is higher than KBi$_2$ as expected.

(d) Melting temperature

The melting temperature ($T_m$) is a significant factor that defines the temperature range over which a solid can be used. Higher cohesive energy, higher bonding energy, and a lower coefficient of thermal expansion are all characteristics of a solid with a higher melting temperature [100]. It is directly related to the bonding strength. At temperatures below $T_m$, solids can be used continuously without oxidation, chemical change, and excessive distortion causing mechanical failure. The elastic constants can be used to calculate the melting temperature $T_m$ of solids using the following equation [100]:

$$T_m = \left[553\text{K} + \left(5.91 \frac{\text{K}}{\text{GPa}}\right) C_{11}\right] \pm 300\ K \qquad (47)$$

The melting temperature of RbBi$_2$ is higher than KBi$_2$ as shown in Table 10.

(e) Minimum thermal conductivity

A limit of a fundamental thermal characteristic is the minimum thermal conductivity. The minimum thermal conductivity ($k_{min}$) is reached at high temperatures above the Debye temperature, is a minimal value for a compound's thermal conductivity. The fact that the minimal thermal conductivity is unaffected by crystal imperfections (such as dislocations, individual vacancies, and long-range strain fields associated with impurity inclusions and dislocations) is crucial. This is due to the fact that at high temperatures, the phonon mean free path dramatically shrinks below the length scale for atomic separation. Clarke derived the following formula to determine the lowest thermal conductivity, $k_{min}$, based on the Debye model of compounds at high temperatures [101]:

$$k_{min} = k_B \upsilon_m (V_{atomic})^{-2/3} \qquad (48)$$

In this equation, $k_B$ is the Boltzmann constant, $\upsilon_m$ is the average sound velocity and $V_{atomic}$ is the cell volume per atom.

The calculated values of minimum thermal conductivity for $XBi_2$ ($X$ = K, Rb) are enlisted in Table 9. $k_{min}$ of $RbBi_2$ is higher than that of $KBi_2$. Compounds with higher sound velocity and Debye temperature have higher minimum thermal conductivity.

The anisotropic minimum thermal conductivity is also present in materials with elastic anisotropy. The lowest thermal conductivity varies with the sound velocity in different crystallographic orientations and is anisotropic. Using the Cahill and Clarke model, the minimum thermal conductivities along various directions are calculated [102]:

$$k_{min} = \frac{k_B}{2.48} n^{2/3} (\upsilon_l + \upsilon_{t1} + \upsilon_{t2}) \qquad (49)$$

$$n = N/V \qquad (50)$$

where, $k_B$ is the Boltzmann constant, $n$ is the number of atoms per unit volume and $N$ is the total number of atoms in the cell having a volume $V$. The minimum thermal conductivity of $XBi_2$ ($X$ = K, Rb) along [001], [110] and [111] directions are summarized in Table 9. Minimum thermal conductivity of $RbBi_2$ is higher than that of $KBi_2$. For $XBi_2$ ($X$ = K, Rb) compounds, the isotropic minimum thermal conductivity is lower than the minimum thermal conductivities along various crystallographic axes.

**Table 9:** Grüneisen parameter $\gamma$, The number of atoms per mole of the compound $n$ (m$^{-3}$), lattice thermal conductivity $k_{ph}$ (W/m. K) at 300 K, and minimum thermal conductivity $k_{min}$ (W/m. K) of $XBi_2$ ($X$ = K, Rb) compounds along different directions.

| Compound | $\gamma$ | $n$ ($10^{28}$) | $k_{ph}$ | $[001]k_{min}$ | $[110]k_{min}$ | $[111]k_{min}$ | $k_{min}$ |
|---|---|---|---|---|---|---|---|
| $KBi_2$ | 2.75 | 2.78 | 0.66 | 0.21 | 0.22 | 0.21 | 0.13 |
| $RbBi_2$ | 1.60 | 2.66 | 0.89 | 0.28 | 0.31 | 0.28 | 0.20 |

(f) Thermal expansion coefficient

The thermal expansion coefficient ($\alpha$) estimates the thermal strain per unit change in temperature. The relationship between a material's thermal expansion coefficient (TEC) and other physical characteristics can be complex. Examples include the correlation between TEC and specific heat, thermal conductivity, temperature-dependent energy band gap, and electron effective mass. When utilized in spintronic and electrical devices, the thermal expansion coefficient ($\alpha$), an intrinsic thermal feature of a material, is crucial for the epitaxial growth of crystals and the mitigation of harmful consequences. The study of a material's thermal expansion coefficient is also useful for assessing a system's potential as a thermal barrier layer. A material's low value of $\alpha$ suggests that it is stable throughout a wide temperature range. The following equation can be used to calculate a material's thermal expansion coefficient [102,103]:

$$\alpha = \frac{1.6 \times 10^{-3}}{G} \quad (51)$$

where, $G$ is the shear modulus (in GPa). The thermal expansion coefficient is inversely related to melting temperature: $\alpha \approx 0.02/T_m$ [101,103]. The computed thermal expansion coefficients of $X\text{Bi}_2$ ($X$ = K, Rb) compounds are disclosed in Table 10.

(g) Heat capacity and dominant phonon wavelength

The heat capacity ($C_P$), a physical property, determines the amount of heat required to change an object's temperature by one degree. Thermal energy change in a material per degree per unit volume is defined by its heat capacity per unit volume ($\rho C_P$). The heat capacity of a material per unit volume is calculated from [102]:

$$\rho C_P = \frac{3k_B}{\Omega} \quad (52)$$

where, $N = 1/\Omega$ is the number of atoms per unit volume. The heat capacity per unit volume of $X\text{Bi}_2$ ($X$ = K, Rb) compounds are given in Table 10. Compared to $\text{RbBi}_2$, $\text{KBi}_2$ has a greater heat capacity per unit volume.

Phonons are crucial in defining a number of physical characteristics, including heat capacity, thermal conductivity, and electrical conductivity. The dominant phonon's wavelength, $\lambda_{dom}$, is the wavelength at which the phonon distribution function its maximum. It is to be noticed that high values of the dominant phonon wavelength results from low density, high shear modulus, and high mean sound velocity. We have calculated $\lambda_{dom}$ for $X\text{Bi}_2$ ($X$ = K, Rb) at 300 K using the following relationship [49,101]:

$$\lambda_{dom} = \frac{12.566 \, v_m}{T} \times 10^{-12} \quad (53)$$

where, $v_m$ is the average sound velocity in ms$^{-1}$, $T$ is the temperature in K. The estimated values of $\lambda_{dom}$ in meters are also listed in Table 10.

**Table 10:** Calculated melting temperature $T_m$ (K), thermal expansion coefficient $\alpha$ (K$^{-1}$), heat capacity per unit volume $\rho C_\rho$ (JK$^{-1}$m$^{-3}$), and wavelength of the dominant phonon mode $\lambda_{dom}$ (m) at 300 K of $X\text{Bi}_2$ ($X$ = K, Rb) compounds.

| Compound | $T_m(\pm 300$ K$)$ | $\alpha$ (10$^{-5}$) | $\rho C_\rho$ (10$^6$) | $\lambda_{dom}$ (10$^{-12}$) |
|---|---|---|---|---|
| KBi$_2$ | 745.28 | 28.8 | 1.15 | 42.25 |
| RbBi$_2$ | 853.99 | 9.56 | 1.10 | 69.51 |

## 4. Conclusions

This research presents a detailed first-principles investigation of the $X$Bi$_2$ ($X$ = K, Rb) compounds in the cubic Laves phase employing the density functional theory. The majority of the results reported herein are novel. In cases where they were accessible, we also compared our findings with previous results and fair agreements were found.

The $X$Bi$_2$ ($X$ = K, Rb) compounds are elastically stable and possess moderate mechanical anisotropy. KBi$_2$ is relatively more anisotropic than RbBi$_2$. Both the compounds are soft, ductile and machinable. The level of ductility and machinability is particularly high for KBi$_2$. It is clear based on the Kleinman parameter values that the main source of mechanical strength in these compounds is the contribution of bond stretching or contracting. Calculations of the electronic energy density of states and the electronic band structure suggest clear metallic properties. The electronic stability of RbBi$_2$ is expected to be higher as indicated by the position of the Fermi level close to the center of the pseudogap in the TDOS profile. The Debye temperature, lattice thermal conductivity, melting temperature, and minimum thermal conductivity of the Laves phases of interest are entirely consistent with each other. Both the compounds have very low Debye temperature and phonon thermal conductivity. These compounds hold promise to be used as TBC materials at relatively low temperatures. The melting temperature and hardness of $X$Bi$_2$ ($X$ = K, Rb) compounds are also quite low. The effect of anharmonicity is large in KBi$_2$. The real part of the refractive index of KBi$_2$ and RbBi$_2$ has a very high value at low energies in the infrared region. The reflectivity is also very high in the visible region for both KBi$_2$ and RbBi$_2$. The compounds have a great deal of promise for usage as a covering that reflects solar energy with high efficiency.

To summarize, the compounds under investigation have a number of desirable mechanical, thermal, and optoelectronic properties that make them useful for engineering and device applications. Our findings should motivate the researchers to explore these interesting binary Laves systems more thoroughly, both theoretically and experimentally.

## Data availability

The data sets generated and/or analyzed in this study are available from the corresponding author on reasonable request.

## Declaration of interest

The authors declare that they have no known competing financial interests or personal relationships that could have appeared to influence the work reported in this paper.

## CRediT author statement

**Jahid Hassan:** Methodology, Software, Formal analysis, Writing-Original draft. **M. A. Masum:** Methodology, Software, Formal analysis, Writing-Original draft. **S. H. Naqib:** Conceptualization, Supervision, Formal analysis, Validation, Writing- Reviewing and Editing.

# List of references